\documentclass[sigconf,nonacm=true,balance=true]{acmart}

\AtBeginDocument{%
  \providecommand\BibTeX{{%
    \normalfont B\kern-0.5em{\scshape i\kern-0.25em b}\kern-0.8em\TeX}}}

\setcopyright{none}

\usepackage{graphicx,subcaption}
\usepackage[autostyle]{csquotes}
\usepackage{listings}
\usepackage{color}
\definecolor{light-gray}{gray}{0.95}
\definecolor{codegray}{rgb}{0.5,0.5,0.5}
\usepackage{enumitem}
\usepackage{pbalance}
\usepackage{hyperref}
\hypersetup{
pdftitle={JsStories: Improving Social Inclusion in Computer Science Education Through Interactive Stories},
pdfsubject={Story-based Learning},
pdfauthor={Inas Ghazouani Ghailani, Yoshi Malaise and Beat Signer},
pdfkeywords={Story-based Learning, Computer Science Education, Social Inclusion, PRIMM Principles, Learning JavaScript}
}

\usepackage{balance}

\newcommand{\code}[1]{\texttt{#1}}

\lstdefinestyle{mystyle}{
  backgroundcolor=\color{backcolour}, commentstyle=\color{codegreen},
  keywordstyle=\color{magenta},
  numberstyle=\tiny\color{codegray},
  stringstyle=\color{codepurple},
  basicstyle=\ttfamily\footnotesize,
  breakatwhitespace=false,         
  breaklines=true,                 
  captionpos=b,                    
  keepspaces=true,                 
  numbers=left,                    
  numbersep=5pt,                  
  showspaces=false,                
  showstringspaces=false,
  showtabs=false,                  
  tabsize=2
}

\AtBeginDocument{%
  \providecommand\BibTeX{{%
    Bib\TeX}}}

\newcommand{\dquote}[1]{``#1''}
\newcommand{\squote}[1]{`#1'}

\begin{document}

\title{JsStories: Improving Social Inclusion in Computer Science Education Through Interactive Stories} 

\author{Inas Ghazouani Ghailani}
\authornote{Both authors contributed equally to this research.}
\orcid{0009-0004-6144-9108}
\affiliation{%
  \institution{WISE Lab}
  \department{Vrije Universiteit Brussel}
  \city{Brussels}
  \postcode{1050}
  \country{Belgium}
}
\email{inas.ghazouani.ghailani@vub.be}

\author{Yoshi Malaise}
\authornotemark[1]
\orcid{0000-0002-3228-6790}
\affiliation{%
  \institution{WISE Lab}
  \department{Vrije Universiteit Brussel}
  \city{Brussels}
  \postcode{1050}
  \country{Belgium}
}
\email{ymalaise@vub.be}

\author{Beat Signer}
\orcid{0000-0001-9916-0837}
\affiliation{%
  \institution{WISE Lab}
  \department{Vrije Universiteit Brussel}
  \city{Brussels}
  \postcode{1050}
  \country{Belgium}
}
\email{bsigner@vub.be}

\begin{abstract}
A main challenge faced by non-profit organisations providing computer science education to under-represented groups are the high drop-out rates. This issue arises from various factors affecting both students and teachers, such as the one-size-fits-all approach of many lessons. Enhancing social inclusion in the learning process could help reduce these drop-out rates. We present JsStories, a tool designed to help students learn JavaScript through interactive stories. The development of JsStories has been informed by existing literature on storytelling for inclusion and insights gained from a visit to HackYourFuture Belgium (HYFBE), a non-profit organisation that teaches web development to refugees and migrants. 
To lower barriers to entry and maximise the feeling of connection to the story, we incorporated narratives from HYFBE alumni. Further, we adhered to educational best practices by applying the PRIMM~principles and offering level-appropriate content based on knowledge graphs. JsStories has been demonstrated, evaluated and communicated to the different stakeholders through interviews and a survey, enabling us to identify future directions for story-based learning solutions.
\end{abstract}

\ccsdesc[500]{Applied computing~Interactive learning environments}
\ccsdesc[500]{Applied computing~Distance learning}

\keywords{Story-based learning, computer science education, social inclusion, PRIMM principles, learning JavaScript}

\maketitle

\section{Introduction}
Learning and teaching programming is known to be difficult~\cite{jenkins2002difficulty, robins2003learning, cheah2020factors}, as evidenced by the high failure rates in programming courses, indicating room for improvement~\cite{watson2014failure}. Several crucial aspects that could be improved in the current approach to teaching programming have been identified~\cite{jenkins2002difficulty}. Suggestions for improvement include making the teaching of programming more flexible, allowing students to learn at their own pace and through different methods such as solitary learning, dynamic group discussions or understanding first before writing. Students should also receive adequate and appropriate assistance. Additionally, it is essential that teachers are not only skilled programmers, but also possess strong pedagogical skills to effectively teach programming.

With the increasing digitalisation, STEM fields such as computer science are growing in importance and demand. However, statistical evidence shows that individuals from less-represented ethnicities and socially vulnerable groups, such as refugees, are even more underrepresented in these fields~\cite{ nassar2011new}. Note that the same applies to people of diverse sexual orientations and gender identities~\cite{freeman2020measuring}.
To lower the entry threshold for these groups, numerous non-profit organisations worldwide aim to address the inequality and high dropout rates among underrepresented groups. For example, several organisations have joined forces in the Migracode Europe project\footnote{\url{https://migracode.eu/about-migracode/}}, co-funded by the Erasmus+ programme of the European Union, which promotes open access technology education for refugees and migrants. 

It is crucial to consider the background and needs of these refugees and migrants to foster their sense of belonging. Addressing these aspects is essential for reducing dropout rates in organisations that aim to teach programming to socially vulnerable groups. Therefore, we believe that solutions should incorporate these considerations alongside the previously mentioned suggestions for improving programming education. In the research described in this paper, we applied the \emph{Design Science Research Methodology (DSRM)} for information systems research~\cite{peffers2007design} to develop an initial JsStories prototype This prototype was then demonstrated to multiple stakeholders in the field to gain a deeper understanding of the problem space and derive future research directions.
\section{Related Work}

\begin{figure*}[htb]
    \centering \includegraphics[width=0.93\linewidth]{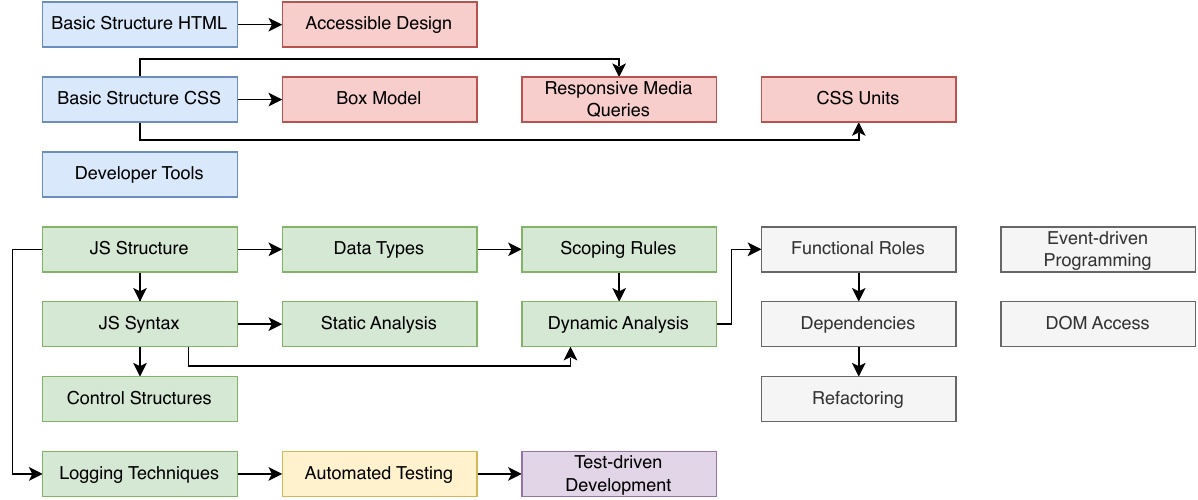}
    \caption{Topics covered by JsStories with colours matching the module of the HYFBE curriculum where they are introduced}
    \label{fig:enter-label}
\end{figure*}

The use of storytelling in computing education has a long history. Almost two decades ago, Kelleher and Pausch explored the idea of incorporating storytelling into introductory programming classes through the \emph{Storytelling Alice} tool. In their work, they found that students experienced the lessons less frustrating and more engaging~\cite{kelleher2007using}. The solution builds on earlier work by Moskal~et~al.~\cite{moskal2004evaluating}, who demonstrated that using an earlier version of \emph{Alice} to create interactive 3D~environments resulted in higher retention rates, particularly for so-called high-risk students with limited background in computer science and mathematics. An interesting insight from Kelleher and Pausch's work is that the storytelling edition of Alice was able to attract individuals who did not initially intend to study computer science. This was not observed with the earlier non-story-focused version of \emph{Alice}, highlighting the power of stories to engage people through intrinsic motivation. 

Research further shows that the frequency and portrayal of social identities in educational materials in the United States affect how and what students learn. In their 2021 research overview \dquote{The Representation of Social Groups in US Educational Materials and Why It Matters}, the author synthesised the results of more than 160~studies and emphasised the need for educational materials to create a sense of belonging for students, affirming their place in learning environments and communities~\cite{armstrong2021representation}.

In addition to the content itself and how it is presented, the order in which elements are shown to students is crucial. 
Bloom's taxonomy~\cite{bloom1956handbook} is a widely adopted model in education with a six-level hierarchical classification of learning objectives. It consists of the \emph{knowledge}, \emph{comprehension}, \emph{application}, \emph{analysis}, \emph{synthesis} and \emph{evaluation} levels, with the latter levels requiring more cognitive load than the earlier models~\cite{forehand2010bloom, adams2015bloom}. Teachers can use Bloom's taxonomy to structure their lessons so that students progress through each level sequentially, only advancing to higher levels once they have mastered the lower ones~\cite{forehand2010bloom}. Building on the same ideas, the Block model was later introduced by Schulte~\cite{block} specifically for programming education. Lee~et~al.~\cite{lee2011computational} proposed the Use-Modify-Create model to help students develop their computational thinking skills. The model is based on the idea that students should first learn how to read and use code before starting to adapt it to their needs, ensuring they have a thorough understanding of a program before having to write any code from scratch. Later, Sentance~et~al.~\cite{primmTeachers, primmStudents} introduced PRIMM, which further divides the lesson structure into the five stages of \emph{p}redict, \emph{r}un, \emph{i}nvestigate, \emph{m}odify and \emph{m}ake.

\section{Design and Development of JsStories}
\label{sec:design-and-dev}
In the following, we provide some details about the design and development stage of the JsStories prototype, following the Design Science Research Methodology~\cite{peffers2007design}. The envisioned artefact was a set of digital stories interwoven with programming exercises that learners could work through, hopefully recognise parts of themselves, and feel validated throughout their learning journey. The stories were collected from HYFBE~alumni who volunteered to share their experiences. The volunteers received an introduction to our research and a clear explanation of how their stories would be used. They were then encouraged to talk freely about their lived experiences, sharing what was most relevant to them. To ensure accurate understanding and avoid any misinterpretations, the \dquote{Listen, Sum Up and Don't Stop Asking Questions}~(LSD) method~\cite{LSD} was used during the interviews. Every interview was recorded with the participants' permission to support later transcription. The notes were summarised in bullet points and converted into narrative text using ChatGPT\footnote{\url{https://chat.openai.com}}. Further, all stories were pseudonymised to maintain the privacy of the participants.

After collecting the stories, we contacted the authors of~\cite{van2023bridging} to learn how they constructed the knowledge graph for their e-learning platform for programming education. Based on insights gained from that conversation---as well as from interviews with HYFBE~students---we analysed the existing HYFBE~curriculum and learning materials to construct the concept graph illustrated in Figure~\ref{fig:enter-label}. We then assigned topics from the concept map to each of the stories, guiding students with a clear progression path while avoiding a strictly linear approach.

Once we had the narratives and identified the topics to focus on in each story, we pinpointed elements or scenes in the story for which we could construct fitting coding exercises. To guide students towards appropriate content, we introduced a locking mechanism that would restrict access to certain stories until the required topics were mastered. To ensure a smooth progression within a story, the incorporated exercises were inspired by the phases of PRIMM as defined by Sentance~et~al.~\cite{primmStudents}. These stages are (1)~Predict, (2)~Run, (3)~Investigate, (4)~Modify and (5)~Make. 

\begin{figure*}[htb]
    \centering
    \includegraphics[width=0.65\textwidth]{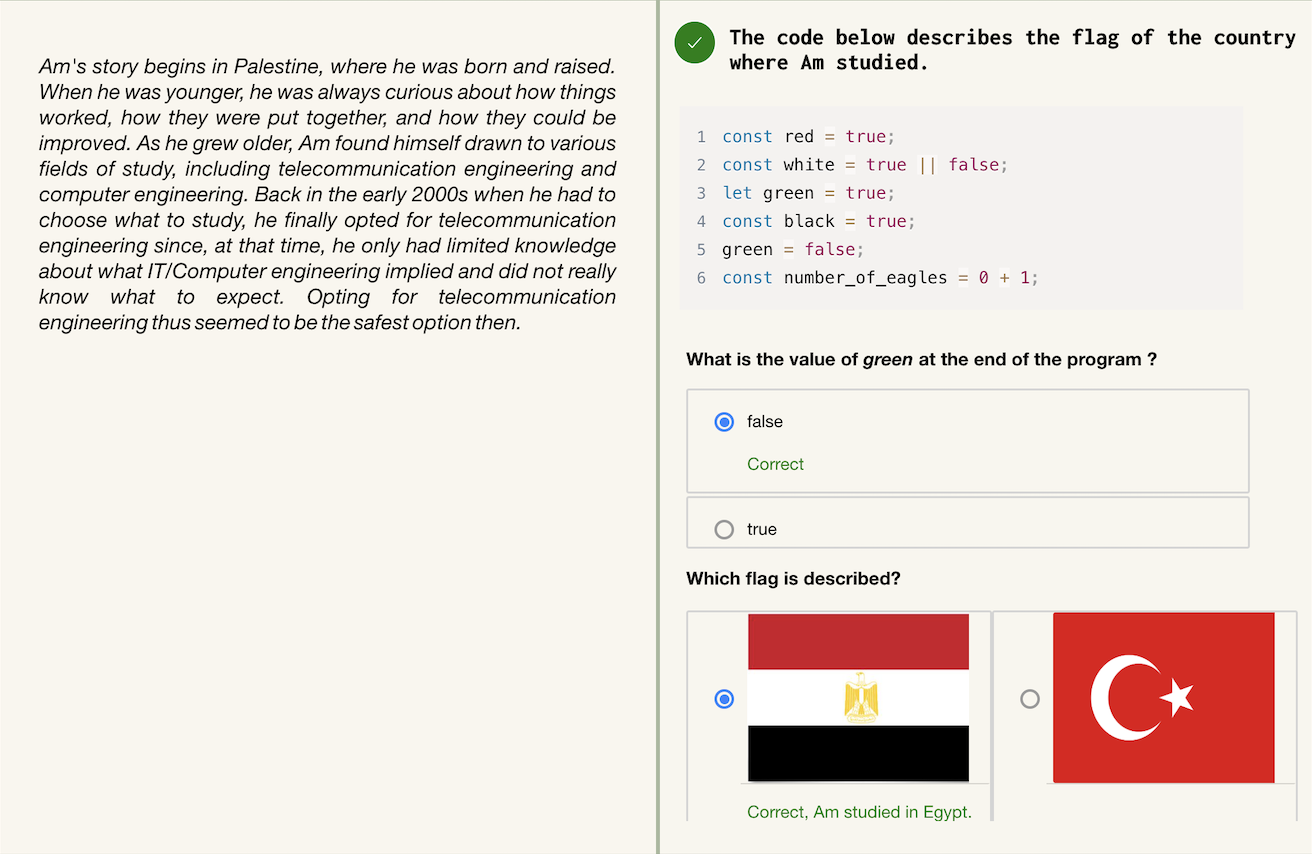}
    \vspace{-0.2cm}
    \caption{Example of a predict exercise}
    \label{fig:predict}
\end{figure*}

For the \emph{predict phase}, we included multiple choice questions that require learners to predict the outcome of code. These questions feature options that may contain both images and text. Figure~\ref{fig:predict} illustrates an example of such a predict exercise. In this particular exercise, the given code describes a flag and the student has to predict which flag it represents from a set of given options. 

\begin{figure*}[htb]
    \centering
    \includegraphics[width=0.65\textwidth]{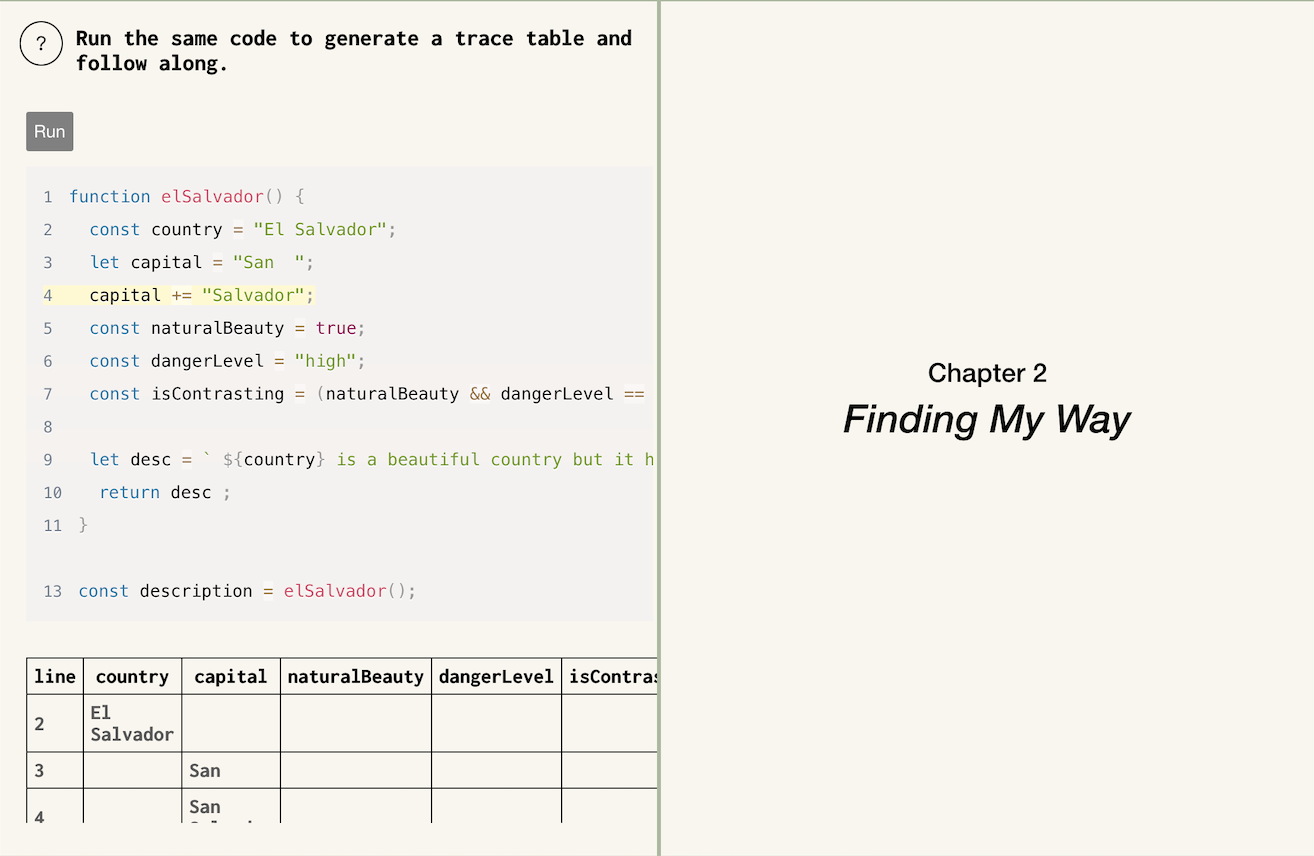}
    \vspace{-0.2cm}
    \caption{Example of a run exercise}
    \label{fig:run}
\end{figure*}

Next, in the \emph{run} phase, students are required to execute the provided code while following along with an automatically filled-in trace table. Each line of the code execution is highlighted, allowing students to observe changes in the trace table for that specific line. This type of exercise was included because tracing code, i.e.~simulating its execution, enhances understanding. An example of this exercise is shown in Figure~\ref{fig:run}. As illustrated, the fourth line of the code currently being executed is highlighted and the trace table shows the value assigned to the \code{capital} variable. Initially, we implemented the run exercise by using the JavaScript parser Acorn\footnote{\url{https://www.npmjs.com/package/acorn}} to generate an abstract syntax tree~(AST). Additionally, \mbox{Acorn-walk}\footnote{\url{https://www.npmjs.com/package/acorn-walk}} was used to traverse the tree and extract all the necessary information to build the trace table. However, this functionality was later reimplemented using the Aran\footnote{\url{https://github.com/lachrist/aran}} library. Aran allows developers to utilise Aspect-oriented Programming to weave advice into the AST of the code. By injecting some code every time a READ or WRITE occurs, we can automatically build a trace table based on any arbitrary piece of JavaScript code.

\begin{figure*}[htb]
    \centering
    \includegraphics[width=0.65\textwidth]{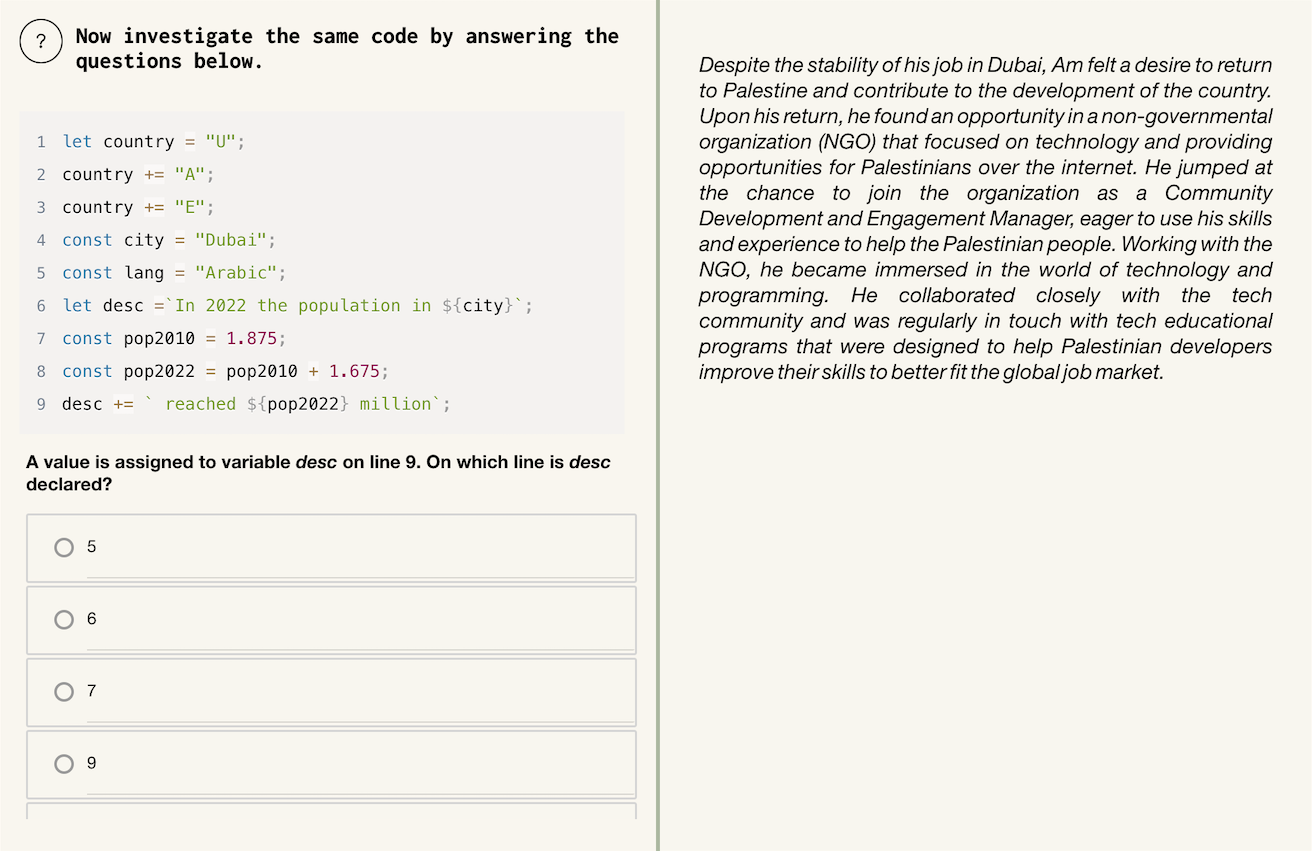}
    \vspace{-0.2cm}
    \caption{Example of an investigate exercise}
    \label{fig:investigate}
\end{figure*}

\begin{figure*}[htb]
    \centering
    \includegraphics[width=0.65\textwidth]{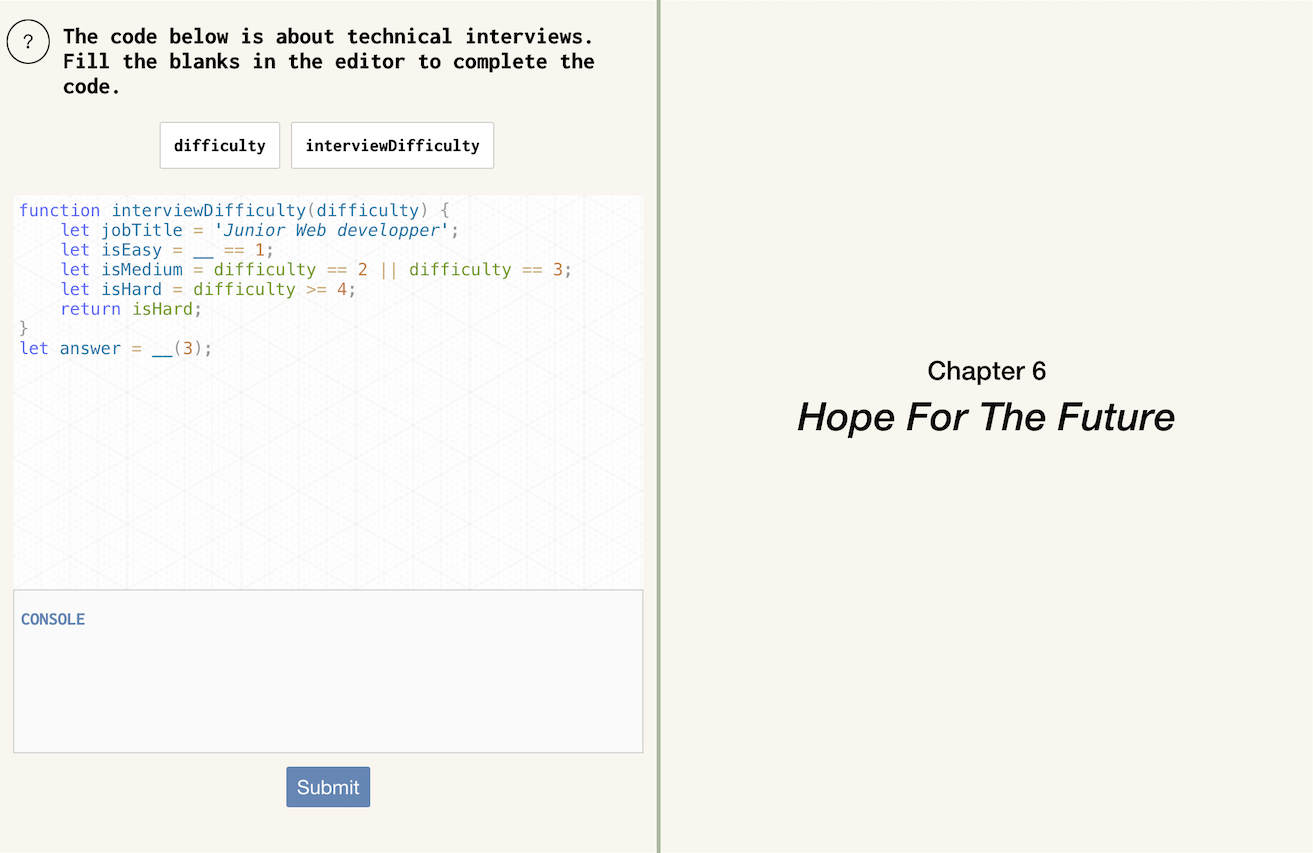}
    \vspace{-0.2cm}
    \caption{Example of a \squote{fill in the blanks} modify exercise}
    \label{fig:blanks}
\end{figure*}

In the \emph{investigate} phase, students focus on understanding code structure. Therefore, exercises involving multiple choice questions that require learners to investigate the structure of code have been included. These questions cover topics such as variable declaration and initialisation as depicted in an example in Figure~\ref{fig:investigate}. The open-source QLC~\cite{lehtinen2021let} library was used to automatically generate between one and three questions about the structure of the code fragment. The library is able to generate seven different types of exercises. Moreover, for each answer option, a brief explanation is provided of why it is or is not the correct answer.

In the \emph{modify} phase, where students finally get to modify code, two types of exercises were incorporated. The first type is a blanks exercise, in which students are given some code containing blanks and options to fill them in. These options can, for instance, include keywords, variable names and values. If a student filled in one of the options in a blank, they can tap the badge to strike it through. An example of this exercise is shown in Figure~\ref{fig:blanks}. To incorporate the blanks, the Acorn parser was used. The code fragment is parsed using Acorn, and Acorn-walk is then used to traverse the tree. When variables and identifiers are encountered, they are potentially replaced by a blank (\textunderscore\textunderscore). The number of generated blanks is based on the difficulty level parameter of the exercise. For the code editor in the user interface, the react-codemirror2\footnote{\url{https://www.npmjs.com/package/react-codemirror2}} library was used.

\begin{figure*}[htb]
    \centering
    \includegraphics[width=0.65\textwidth]{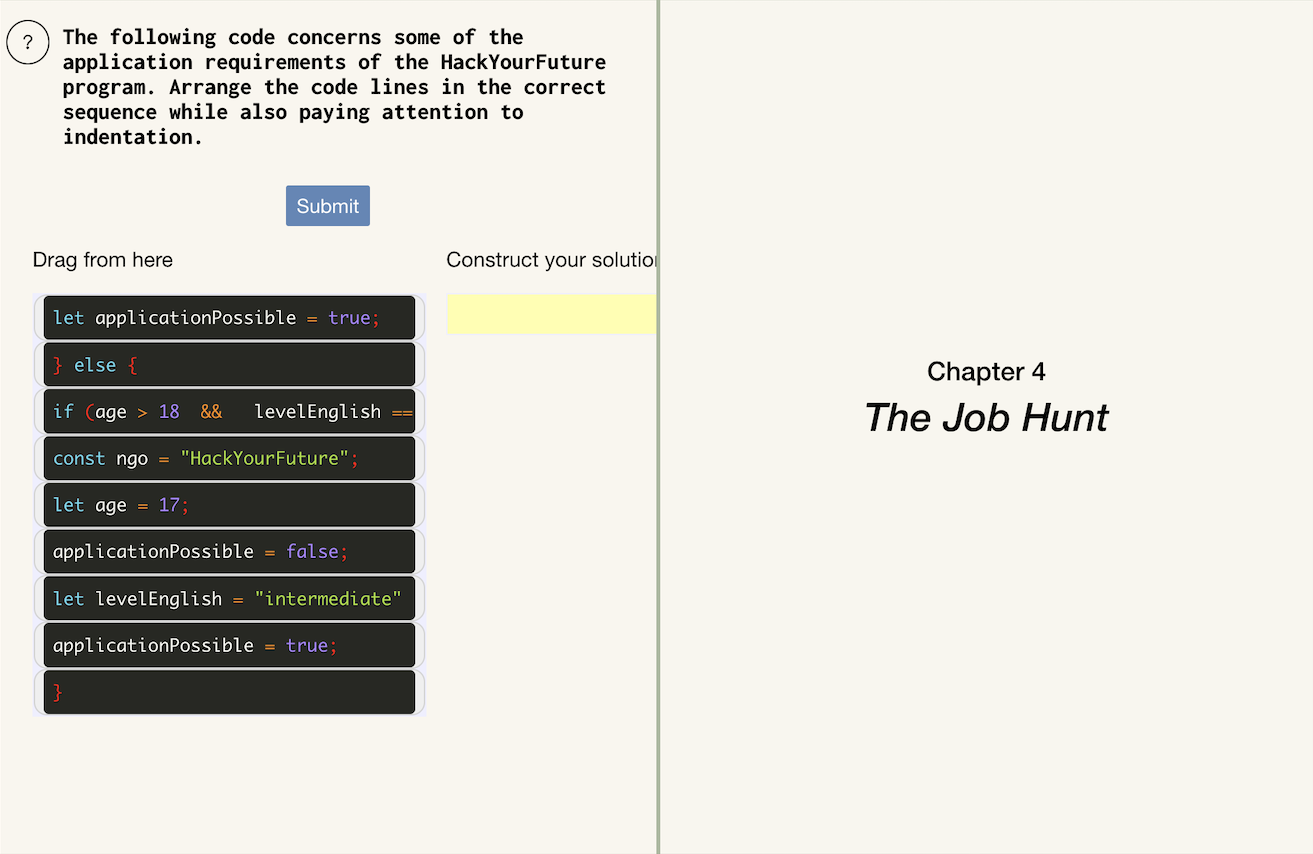}
    \vspace{-0.2cm}    
    \caption{Example of a \squote{Parsons puzzle} modify exercise}
    \label{fig:parsons}
\end{figure*}

\begin{figure*}[htb]
    \centering
    \includegraphics[width=0.65\textwidth]{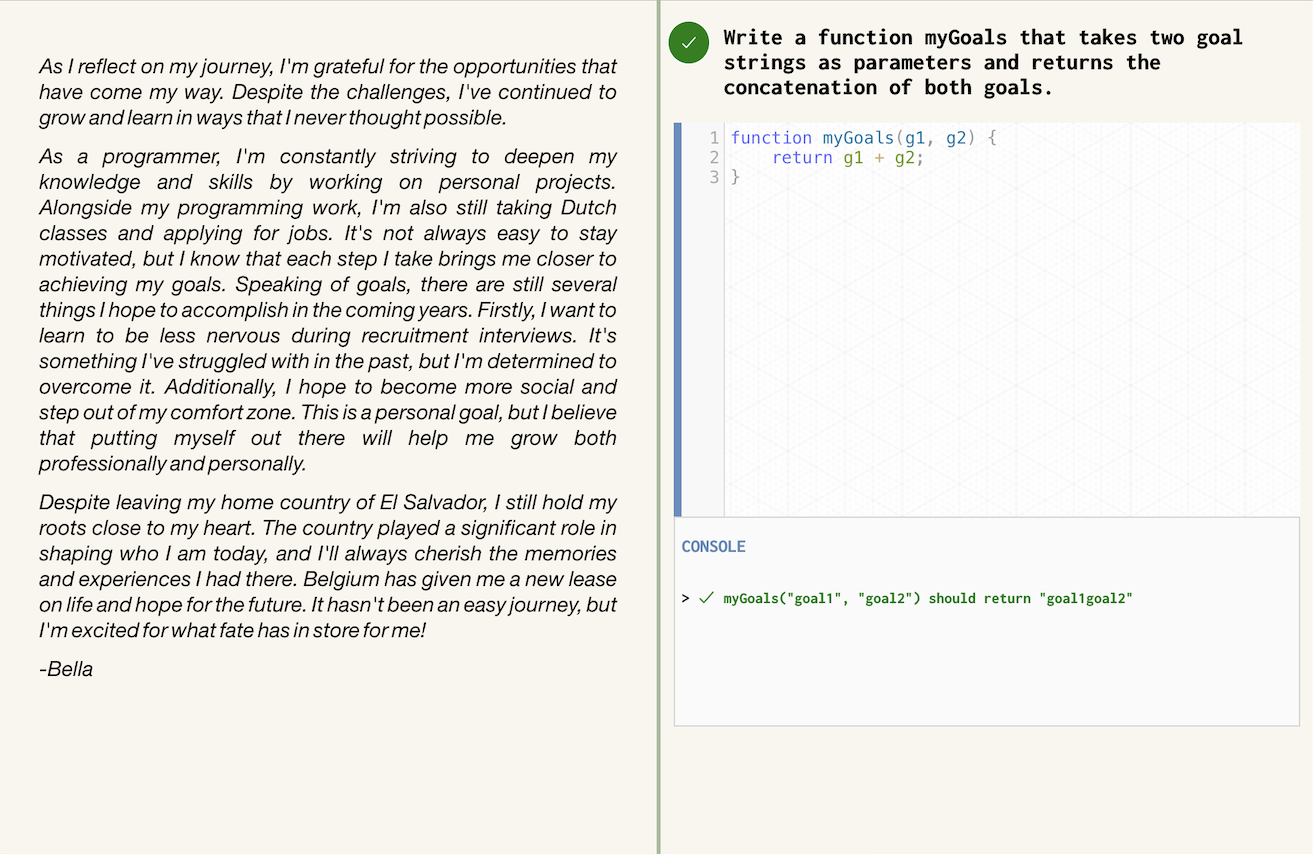}
    \vspace{-0.2cm}    
    \caption{Example of a make exercise}
    \label{fig:make}
\end{figure*}

The second type of exercise is a Parsons puzzle~\cite{parsons2006parson}. In a Parsons puzzle, code is fragmented into several code blocks or lines that appear in the wrong order. The goal is to rearrange these code fragments to form the correct code. Parsons puzzles have been shown to be an effective exercise for learning programming~\cite{denny2008evaluating}. In our implementation, we opted for two-dimensional Parsons puzzles, which require students to put code fragments in the right order while also considering indentation~\cite{ericson2017solving}. Figure~\ref{fig:parsons} illustrates a Parsons puzzle as it appears in the context of a JsStories story. 

Finally, in the exercises for the \emph{make} phase, students are required to write code, which is then checked using predefined unit tests. They are given a description of the required functionality and have to write code accordingly. For instance, in the exercise depicted in Figure~\ref{fig:make}, students are tasked with writing a function with a specific name, number of arguments and return value. The unit tests can be configured by the educator and specific feedback strings can be shown based on which tests are failing.

\begin{figure}[htb]
    \centering
    \includegraphics[width=0.8\columnwidth]{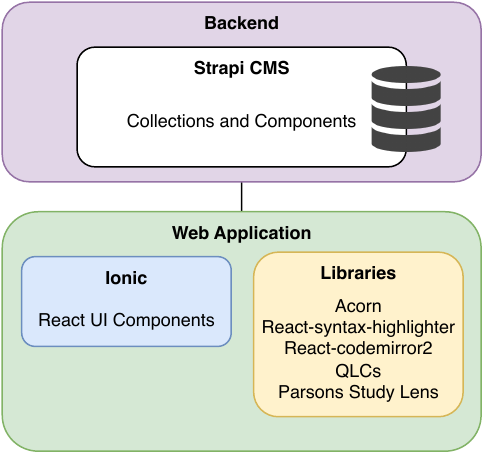}
    \caption{High-level application architecture}
    \label{fig:architecture}
\end{figure}

An implementation of the solution described above was made in React using the Ionic framework\footnote{\url{https://ionicframework.com}} and is available on Github\footnote{\url{https://github.com/InasGhazouani/jsStories}}. Currently, the JsStories tool is available as a desktop application, packaged and distributed using the Electron framework\footnote{\url{https://www.electronjs.org}}. The high-level architecture of the JsStories tool is depicted in Figure~\ref{fig:architecture}. Since the tool is to be used in an educational setting, including teachers and organisational personnel, who may not have a technical background, it should be easy to manage and adapt stories and exercises. Hence, it was decided to separate the content, specifically the stories and exercises, from the more technical aspects of the tool. Therefore, Strapi\footnote{\url{https://strapi.io/}}, an open source headless Content Management System~(CMS), was used. The content can be managed through an admin panel without requiring code changes.

\section{Demonstration and Evaluation}
JsStories has been demonstrated and evaluated by different stakeholders to gain insights about its potential impact in a real-world context. Since the JsStories application is intended for use in educational settings with members of socially vulnerable groups such as refugees and asylum seekers, we sought feedback from students, educators and organisational staff of organisations within this domain. The evaluation process consisted of two stages: first we conducted interviews with stakeholders and then we asked them to share an online survey with their coaches, colleagues and students. 

\subsection{Interviews}
We conducted a total of six interviews with alumni who had previously shared their stories, as well as with members of HYF Belgium, HYF~Denmark, Borders None (Kroatia), Open Cultural Center (Spain) and \mbox{Social} Hackers Academy (Greece). These semi-structured interviews were carried out via the Microsoft Teams platform\footnote{\url{https://www.microsoft.com/nl-be/microsoft-teams/log-in}}. Each individual interview started with the interviewee providing some details about themselves. We then introduced our research and demonstrated the features of JsStories before collecting feedback via a semi-structured interview.

\subsubsection{HYF Belgium} 
As a former education coordinator at HYFBE, the first interviewee's responsibilities primarily involved developing and implementing the curriculum. Prior to that role, he served as an educational officer, assisting the students and closely monitoring their learning journeys. Overall, the interviewee was enthusiastic about the application and appreciated its storytelling aspect. He particularly liked that each story is designed in book form, with integrated exercises that fit the storyline. He believes that the tool has much to offer students who are still doubting whether programming is the right thing for them. The applicability of the tool within the HYFBE or a similar programme was then discussed. In terms of exercises, the interviewee noted that the current version of JsStories contains a limited number of exercises and suggested including optional exercises of the same type and covering the same learning features after each exercise. Additionally, he proposed adding a free practice environment at the end of the story with a summarising table of contents. This table of contents would summarise all the learning features covered in the story, along with external links and references. 

When asked about his opinion on using JsStories as part of the selection procedure for a programme such as HYF~Belgium, the interviewee liked the idea. He appreciated the fact that in JsStories the story itself is the content itself rather than a separate additional resource. He also noted that testing someone's ability to write code at the start of a programming course is not helpful, which is why he valued the tool's use of the PRIMM approach instead of simply asking students to write code. Nonetheless, he stated that if the tool is used for entrance examinations, extra support should be provided to prevent students from feeling disoriented. For example, a help button could show a recording of how to complete the exercise. Additionally, since trace tables are assumed to be new for students beginning their programming journey, it should not be expected that they will automatically understand that they have to read the code, follow the automatically filled-in trace table and make the connection between these two. To support his claim, he referred to the expertise reversal effect, stating that novices and experts learn differently, with novices requiring more step-by-step approaches~\cite{schnotz2010reanalyzing}. Therefore, the automatically filled-in trace tables might not be suitable for beginners. Instead, he recommended adapting the run exercises to include blanks that students must fill in, rather than having the trace table filled in completely automatically. Alternatively, multiple choice questions could be included after the trace table. 

The second interview was with an HYFBE~alumnus who contributed by sharing their story. The interviewee was first asked about the potential impact of the tool's storytelling aspect on learners' motivation and willingness to engage with the content and exercises. They believe that it can have a positive effect but noted that it largely depends on the story and how much the learner can relate to it. Additionally, they stated that the quality of the story's writing is crucial. When asked whether such stories could enhance social inclusion, they affirmed that reading happy ending stories of people who have experienced similar situations could bring a sense of hope to the reader. For instance, reading about someone who successfully completed their programming journey despite facing similar challenges may motivate a reader who is uncertain about continuing their programming journey and prevent them from dropping out. 

Moreover, they mentioned that they would have appreciated having access to such a tool when they were following the HYFBE~programme and believe it may be useful for the JavaScript module. They also commented that the tool could serve as a means of preparation for the programme, giving novices an idea of what to expect before starting their programming journey. In addition, they valued the tool's use of the PRIMM methodology, providing step-by-step support for students. When asked about the design of the tool, they mentioned liking the usage of the book-style design, including text, images and exercises. However, they remarked that the design of the Parsons exercise could be improved, as its purpose was not clear at first due to the need for horizontal scrolling. When asked whether they had any other general remarks or feedback, they expressed gratitude for the efforts put into this work to help socially vulnerable people. They found the tool promising and envisioned it being used not only at HYFBE but also in other similar programmes.

\subsubsection{HYF Denmark}
For HYF Denmark\footnote{\url{https://www.hackyourfuture.dk}}, the interview was conducted with the founder of the organisation as well as the current managing director. After demonstrating JsStories, the interviewees had some questions about the tool's functionalities, including whether mentors could monitor the knowledge graph or learning path of the students. Although this feature is not currently available, it would be a useful feature to add in the future. Since the tool is currently based on the HYFBE curriculum, the used knowledge graph was explained and the relevant modules were compared to those of HYF Denmark's curriculum. The interviewees were then questioned about the storytelling aspect and the used stories, specifically whether they think it could help improve student motivation, engagement and social inclusion. They believe that, in general, reading and considering stories of diverse people and experiences might have a positive impact, especially for non-traditional boot camps aiming to teach socially vulnerable people. However, they noted that HYF Denmark's strong focus on getting their students into the job market means that they target highly motivated and passionate students who do not need additional motivation during the programme. The managing director further commented that stories involving real job market cases, working experiences and tips might be more appropriate for HYF Denmark.

When we discussed the possibility of using the tool in the student selection procedure, both interviewees were enthusiastic about it. The current selection procedure at HYF Denmark involves a technical assignment using HTML, CSS and some JavaScript. They noted that for the selection procedure, it is important to find the right balance so that the assignment is challenging enough to assess the skills of the applicant but not overly challenging as to disqualify too many applicants. The primary purpose of this assignment is to test applicants' dedication and time management skills. Therefore, for the technical assignment, freeCodeCamp is utilised as they are familiar with the workload and time needed to solve the exercises. However, one limitation of freeCodeCamp is that the included exercises and projects often reflect a Western-centric perspective. For instance, many exercises feature cats as cute pets, which might seem unfamiliar to people from cultures where cats are not commonly kept as pets. Consequently, they found that JsStories, with the type of stories used, would be well suited for the student selection process. Furthermore, as the interviewees did not have a technical background, we did not question them about the type of exercises selected for the PRIMM approach.

\subsubsection{Borders None}
The fourth interview was conducted with a representative of Borders None\footnote{\url{https://www.bordersnone.com/coders-without-borders/}}, located in Croatia. This organisation focuses on training refugees to pursue a career in web development. Borders None is a smaller organisation with fewer students. The interview was conducted with a member of the directing team at Borders None. Overall, the interviewee expressed enthusiasm for the tool. She characterised it as a solution that \dquote{catches two birds with one stone}, serving as both an exercise tool to enhance students' knowledge and as a motivation booster.

As a social worker, the interviewee regularly uses storytelling and believes it can be useful in numerous situations. In this particular case, she believes the storytelling aspect can boost students' motivation to complete the programme, as well as enhance their sense of belonging. She explained that providing students with real-life stories of people they can relate to may be highly beneficial, as students are interested in whether there are graduates who have completed the course and found employment in the field of web development.

Regarding the use of the JsStories tool in the student selection process, the interviewee explained that at Borders None, the programme is divided into a beginner course and an advanced course. The beginner course covers HTML and CSS, after which students must pass a final project. JavaScript and other more advanced topics are then covered in the advanced course. For the beginner course, students are selected based on discussions with them, assessing their motivation and skills. Therefore, the interviewee believes that the tool is only suitable for use during the course and homework sessions but not in their selection procedure. Furthermore, she mentioned that Borders None has a high dropout rate, mainly because students lose motivation or lack self-confidence as foreigners. Hence, she suggested that an adaptation of the tool covering HTML and CSS might be useful in the beginner course to foster initial motivation and a sense of belonging among the students, helping them stay motivated and boosting their chances of completing both courses.

Moreover, as language courses are offered at Borders None, the interviewee commented that an adapted version of the tool aimed at teaching languages using such stories would also be beneficial. She stated that learning from concrete examples one can relate to makes it much easier to focus and remember. Therefore, she always advises teachers at Borders None to make examples as concrete as possible and to link them to the students' situations and preferences. When asked about the design of the tool, she mentioned liking the book form design and found the tool very intuitive.

\subsubsection{Open Cultural Center}
We conducted an interview with the programme manager of Migracode, one of the two open-access code academies at the Open Cultural Center. The interviewee valued the JsStories tool, finding it a promising and interesting learning resource. Regarding the design of the tool, the interviewee liked the book format and the integration of an editor in some of the exercises. Additionally, he appreciated the inclusion of various types of exercises, allowing students to interact with code in multiple ways. The interviewee particularly liked the idea of identifying students' mistakes and recommending optional exercises based on them. He noted that learners generally enjoy having the choice to complete these optional exercises. He mnetioned that without proper testing, it is challenging to determine the tool's effectiveness. However, he stated that including the tool in the programme would likely be beneficial as it provides an additional learning resource. Furthermore, the interviewee highlighted the use of Flexbox Froggy\footnote{\url{https://flexboxfroggy.com}} in their programme on CSS. This gamified learning resource is highly appreciated by students for its engaging approach to learning CSS. Currently, they are exploring ways to incorporate more gamification into their programme. He thinks that the storytelling aspect of the JsStories tool is a powerful feature that can further enhance engagement, as students can relate to the stories. He also mentioned that the tool might improve social belonging among students, providing a first step towards enhancing social inclusion.

The interviewee believes that, depending on how well JsStories aligns with their curriculum, the tool could definitely fit into the programme as a companion resource for students to practice. When asked whether he thinks that the tool could also be used in the student selection procedure, he expressed strong appreciation for the idea. As part of Migracode's selection procedure, students are currently required to complete a course on Khan Academy\footnote{\url{https://www.khanacademy.org}} and develop a small project using HTML and CSS. Although the project does not include JavaScript at the moment, the interviewee mentioned that they are considering its inclusion. He explained that when starting with the JavaScript module, some students feel overwhelmed learning about the programming concepts for the first time, which sometimes leads to drop outs and it would therefore make a lot of sense to use the JsStories in the student selection procedure. It would introduce students to JavaScript before they join the programme and offer them insights into what to expect. When discussing other directions for the tool, the interviewee shared that adapting the tool for their language lab would be beneficial.

\subsubsection{Social Hackers Academy}
The last interview was held with the co-founder and executive director of Social Hackers Academy~(SHA)\footnote{\url{https://socialhackersacademy.org}}. SHA is a non-profit organisation primarily aimed at teaching web development to refugees and other socially vulnerable groups. The interviewee first expressed gratitude for our research aimed at improving the social inclusion of vulnerable groups. He explained that although SHA initially started as an organisation to help refugees and migrants, they have now shifted to also allow less vulnerable people to join the programme by paying a small fee, while more vulnerable individuals receive scholarships. This adjustment was necessary due to the lack of funding, which is not as readily available as in other countries.

Following the \mbox{COVID-19 pandemic}, the SHA programme moved completely online. As a result, they have students from numerous countries, although more than half their students still reside in Greece. When asked about the storytelling aspect and its impact on students, the interviewee indicated being a big fan of storytelling and found it an interesting way to learn. Overall, he found the tool to be a great concept and confirmed that their students also have a lot of difficulties with JavaScript. However, he noted that the tool's usefulness depends on whether the learning objectives can be achieved using it, which can only be determined by testing the tool in a real setting. Furthermore, he stated that it also depends on the learning modalities, as not everyone learns the same way, and the tool might not be effective for all.

\subsection{Survey}
Our survey was conducted using a Google Form\footnote{\url{https://www.google.com/intl/en_be/forms/about/}}, which included a video demonstration of the tool along with various questions, such as multiple choice, linear scale and open-ended questions. The survey was shared in three organisations, including Borders None, HYF Denmark and the Open Cultural Center (specifically Migracode), primarily through the organisations' Slack platforms. We received a total of seven responses, which, unfortunately, is a rather low response rate.

\begin{figure*}[htb]
    \centering
    \includegraphics[width=\textwidth]{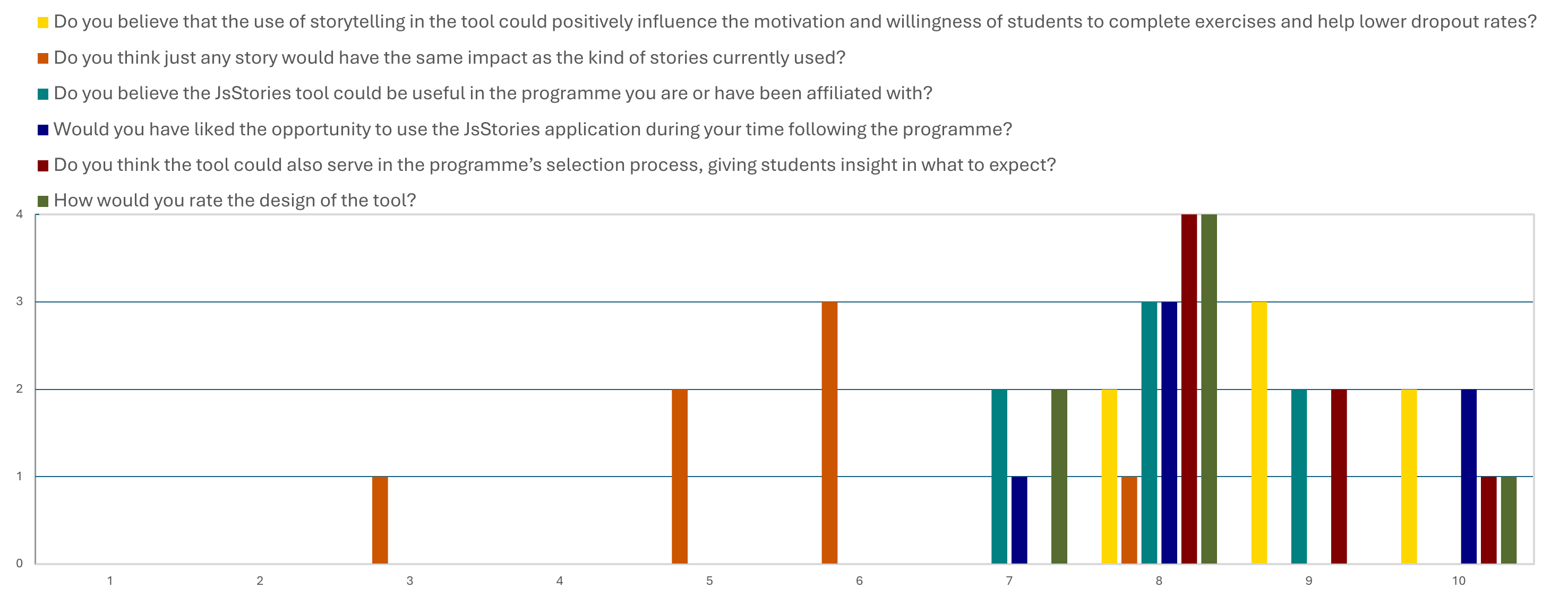}
    \caption{Responses for some of the survey questions on a numerical scale from 1 to 10}
    \label{fig:survey-graphs}
\end{figure*}

The survey was designed with sections and conditional questions to accommodate both students and teachers. One question was specific to teachers, while two questions were tailored for students. Among the seven respondents, one was a former teacher and six were students. When asked whether they believe that the use of storytelling in the tool could positively influence students' the motivation and willingness to complete exercises, potentially reducing dropout rates, all participants responded positively ($M=9$, $SD=0.57$ on a numerical scale from 1 to 10) as illustrated in Figure~\ref{fig:survey-graphs}. The next question, using the same linear scale, focused on the impact of the specific type of stories used. Respondents were asked whether they believe any kind of story would have the same impact as the current real-life stories of HYFBE alumni. Responses were slightly more diverse and overall neutral ($M=5.57$, $SD=1.95$). These answers suggest that a slight majority of the respondents are not convinced that our inclusive stories might have a higher impact than arbitrary stories. Based on these results, we realise that an additional question aimed at assessing the impact of the current stories on students' sense of belonging might have been beneficial.

Respondents were asked if they believe the \mbox{JsStories} tool could be useful in their programme and all participants agreed that it might be beneficial ($M=8$, $SD=0.57$). Based on their responses, the six students who answered this question would have liked the opportunity to use JsStories in their programme ($M=8.5$, $SD=1.25$). The potential use of the tool in the student selection process, providing insights into what to expect, was positively received by all participants ($M=8.57$, $SD=0.53$). The tool's design was also perceived positively ($M=8$, $SD=1$), and all respondents favoured a web version of the tool over a desktop application. A short-answer question regarding the selection of exercises for the PRIMM approach was provided to the teachers. The sole teacher respondent approved the current selection of exercises, finding them appropriate for PRIMM. Students were asked if they would be interested in anonymously contributing their stories for use in the tool. Five out of the six students expressed interest, implying that their stories might be used in future versions of JsStories.

\section{Discussion and Future Directions}
Throughout this work, we communicated the insights gained during the design, development and evaluation of the JsStories artefact, which aims to use storytelling based on alumni's lived experiences interwoven with coding exercises for social inclusion. The obtained feedback indicates that using alumni's real (anonymised) stories might foster a sense of belonging among students and boost their motivation. Based on the students' survey responses, it seems that opinions on whether the social story itself forms an important part of the tool are more divided. However, these results only provide some preliminary findings since we did not get enough responses to make statistically significant conclusions, and there might have been some bias in who decided to answer the survey (i.e.~students participating in a survey posted in Discord groups might be people who feel more comfortable in the environment). Furthermore, the PRIMM approach was appreciated as it provides support for students by guiding them step by step and allowing them to interact with code in different ways. There also seems to be a need for some organisations to include more tools to monitor the learner's progress and gain insights. In our current JsStories prototype, progress is modelled in the form of a personal knowledge graph, but in future work, this could definitely be expanded by evaluating multiple techniques. When deciding upon these techniques, it will be important to keep the structure and mode of operation of the individual organisations in mind. Some organisations tend to be quite centralised, with one or more pedagogical persons responsible for following the entire journey of the students. In such scenarios, a more traditional Learning Management Systems approach, where educators can log in to assess the progress of individual students, might be beneficial. However, some organisations try to model themselves in a more decentralised way, where the goal is that all educational content should be learnable by anyone interested without necessarily being tied to a classroom setup and anyone should be able to host a study group or class group locally. Often, these types of organisations are also more dependent on volunteer coders who help out over short periods of two to three weeks rather than for the entire duration of a course, or they might even just be available on communication channels such as Slack or Discord. In this context, the central platform approach to track progress might be less ideal, and it might be more worthwhile to investigate ways to help students improve their own self-efficacy and direction. Making it easier for them to self-assess gaps in their knowledge and skills so they know specific things to reach out to those volunteers might be a better aid than expecting short-term volunteers to screen all potential remote students for blind spots in their knowledge.

The community also needs a better tool to perform student selection without unnecessarily filtering out candidates. However, what exactly makes for a good entrance test is a much-discussed problem with many different suggested solutions. In general, organisations aim for it to be possible to enter without any prior coding experience. However, some still include JavaScript coding exercises as part of the entrance exam to verify whether students are capable of working independently and searching for solutions in the face of technical challenges to see if they might be a good fit. Some organisations try to assess similar skills by focusing more on Markdown or HTML. While also being technical and potentially leading to errors that need to be debugged, the barrier to entry tends to be lower than the JavaScript exercises. A Markdown-focused version of JsStories might be a promising tool to investigate in the future. Both approaches to assessing which students to accept have advantages and disadvantages, and we believe there is definitely a need for thorough research to investigate this specific question.

The next steps for our project include setting up a more thorough evaluation with real students using JsStories over an extended period to measure the actual effect on their learning journey. We might also consider collaborating with some of the organisations to explore how the environment could be adapted to the language learning domain, as multiple organisations we spoke to mentioned that there is a good fit with this domain.

\begin{acks}
We would like to thank everyone who contributed by sharing their stories, feedback and insights. Special thanks to Laura van der Lubbe for her invaluable expertise in helping us construct the knowledge graph.
\end{acks}

\bibliographystyle{ACM-Reference-Format}
\bibliography{TR-WISE-2025-02}

\end{document}